\documentclass[journal]{IEEEtran}
\usepackage{amsmath,amsfonts}
\usepackage{algorithmic}
\usepackage{algorithm}
\usepackage{array}
\usepackage[caption=false,font=normalsize,labelfont=sf,textfont=sf]{subfig}
\usepackage{textcomp}
\usepackage{stfloats}
\usepackage{url}
\usepackage{pdfpages}
\usepackage{verbatim}
\usepackage{graphicx}
\usepackage{balance} 
\usepackage{cite}
\usepackage{tabulary}
\usepackage{multirow}
\usepackage{textcomp}
\usepackage{tabularx}
\usepackage[T1]{fontenc}
\usepackage[export]{adjustbox}
\newcolumntype{M}[1]{>{\centering\arraybackslash}m{#1}}
\newcolumntype{R}[1]{>{\raggedleft\arraybackslash}m{#1}}
\newcolumntype{L}[1]{>{\raggedright\arraybackslash}m{#1}}
\usepackage{xcolor}

\usepackage{pifont}% http://ctan.org/pkg/pifont
\newcommand{\cmark}{\ding{51}}%
\usepackage{tcolorbox}
\usepackage{enumitem}
\usepackage{graphicx}
\usepackage{booktabs}
\newtcolorbox{suggestion}{breakable,colback=pink!30}
\begin{document}

\sloppy

\title{Appraisal Dimensions Generalise Better than Emotion Labels for Cross-Age Affect Recognition in AI-Assisted Healthcare}

\author{Hippolyte Fournier, Safaa Azzakhnini, Sina Alisamir, Isabella Zsoldos, Eléonore Trân, Gérard Bailly, \\Frédéric Elisei, Béatrice Bouchot, Brice Varini, Patrick Constant, Joan Fruitet, Franck Tarpin-Bernard,\\Solange Rossato, François Portet, Olivier Koenig, Hanna Chainay, Fabien Ringeval%~\IEEEmembership{Staff,~IEEE,}

\thanks{
\textit{
H. Fournier, S. Azzakhnini, S. Alisamir, F. Portet, S. Rossato, and F. Ringeval are with the Univ. Grenoble Alpes, Inria, CNRS, Grenoble INP, LIG. E-mail:\{hippolyte.fournier, sina.alisamir, safae.azzakhnini, francois.portet, solange.rossato, fabien.ringeval\}@univ-grenoble-alpes.fr.\\
H. Chainay, O. Koenig, I. Zsoldos, and E. Trân are with the Univ. Lyon 2, EMC. E-mail:\{hanna.chainay, olivier.koenig, isabella.zsoldos, eleonore.tran\}@univ-lyon2.fr.\\
G. Bailly and F. Elisei are with the GIPSA-lab, Univ. Grenoble Alpes. E-mail:\{gerard.bailly,frederic.elisei\}@gipsa-lab.grenoble-inp.fr.\\
B. Bouchot and B. Varini are with ATOS company. E-mail:\{beatrice.bouchot,brice.varini\}@atos.net.\\
P. Constant is with Pertimm company. E-mail:\{patrick.constant\}@pertimm.com.\\
J. Fruitet and F. Tarpin-Bernard are with Humans Matter company. E-mail:\{j.fruitet, f.tarpin\}@humansmatter.co.\\}

This work has been submitted to the IEEE for possible publication. Copyright may be transferred without notice, after which this version may no longer be accessible.

}
}

\maketitle

\begin{abstract}
Artificial intelligence is increasingly used in healthcare interventions, yet robust affect recognition remains challenging, particularly when models must generalize across patient populations. In this work, we extend the THERADIA-WoZ corpus of AI-assisted Computerized Cognitive Training (CCT) interactions with a new dataset collected from young adults, enabling a direct comparison of affect recognition across age groups. We investigate whether multimodal models based on appraisal dimensions provide a more robust representation of affect than categorical emotion labels. Models were evaluated using within-corpus, cross-corpus, and mixed-corpus training strategies. Results show that appraisal dimensions consistently outperform categorical labels across all evaluation settings. While label-based models fail to generalize across age groups, dropping to chance-level performance in cross-corpus evaluation, appraisal-based models maintain predictive performance above chance. Mixed-corpus training does not further improve generalization beyond within-corpus training. These findings support appraisal dimensions as a more stable and generalizable representation for multimodal affect recognition in healthcare and highlight their potential for AI-assisted interventions involving diverse populations.

\end{abstract}

\begin{IEEEkeywords}
Appraisal theories, Dimensional / Categorical Affect Recognition, Generalisation, Ecological corpus, Computerised Cognitive Training
\end{IEEEkeywords}

\section{Introduction}
\label{sec: intro}
\IEEEPARstart{T}{he} use of artificial intelligence (AI) technologies in healthcare has expanded considerably in recent years. Despite these advances, enabling AI systems to effectively manage social interactions remains a significant challenge \cite{YU18-AI}. Yet, managing social interactions, defined as the ability of autonomous agents to maintain equilibrium in a dynamic relationship with another agent \cite{DEJAEGHER10-SOCIALSKILL}, is particularly important in healthcare contexts. For example, delegating social interactions to AI systems acting as coaches in Computerised Cognitive Training (CCT) could help reduce clinician workload \cite{HILL17-CCT}. CCT consists of a set of exercises designed to train specific cognitive functions. It is primarily used to enhance or preserve cognitive abilities in patients with cognitive impairment. However, the effectiveness of CCT appears to depend on the presence of social interaction during sessions, which currently requires the involvement of a clinician \cite{LAMPIT14-CCT}.

One of the primary challenges in developing virtual assistants for CCT is the difficulty AI systems face in accurately detecting patients' affective states \cite{pepa2021automatic}. To address this limitation, the THERADIA-WoZ corpus was developed \cite{fournier2025theradia}, containing annotated audiovisual segments of older adults (healthy and with mild cognitive impairment) performing CCT sessions assisted by a virtual assistant operated via a Wizard-of-Oz (WoZ) paradigm.

CCT is not only relevant for older adults suffering from age-related cognitive decline; it is also increasingly applied to younger populations, particularly in the context of supporting individuals with learning difficulties such as dyslexia or dyspraxia~\cite{BASHARPOOR24DYS}. These interventions could also benefit from AI-driven virtual assistants capable of providing socially and emotionally adaptive interactions. This raises a central question at the crossroads of cognitive psychology and affective computing: can affect recognition models trained on one age group (e.g., older adults) generalise effectively to another (e.g., younger adults)? A growing body of literature suggests that such generalisation is challenging, as affect appear to be expressed differently across age groups—for example, in terms of facial expressions~\cite{GAYA25DEEP, FOLSTER14EXP, KO21CHANGE, SONMEZ19COMP, ATALLAH19REV, JANNAT21EXP, PARK22FAC}, prosody, and text~\cite{AMORIM21CHAN, LIN24AGE, SEN18AGE, ANDY22TWIT}.

However, the majority of these studies use categorical representations of affect through discrete labels (e.g., fear, anger). Part of affect theorists suggest that labels, as social constructs, may fail to fully capture the underlying processes driving affective expression \cite{BARRETT06SOL}. Instead, they propose that affect refers to changing states representable by dimensions, whose labels would try to capture certain spaces. The most popular dimensional theory proposes to differentiate affect based on valence, ranging from unpleasant to pleasant, and arousal, ranging from deactivation to activation \cite{RUSSEL80-CIRCUM}. While this approach offers greater flexibility than categorical labels, it suffers from limitations as well. Certain labels such as anger and fear occupy the same position in the circumplex, and valence is itself a multidimensional concept \cite{FOURNIER22-COMB, fournier2024emotional}. It has therefore been proposed that two dimensions are not sufficient to differentiate affective states \cite{FONTAINE07-DIM}.

Appraisal theories present a robust dimensional framework to overcome these criticisms \cite{SANDER13-MOD}. These theories posit that affective states arise from an individual's cognitive appraisal of an event's significance, based on a structured set of criteria covering the event's relevance to well-being, its implications, the individual's ability to cope, and its alignment with personal or social norms \cite{SANDER05-CPM}. This structured evaluation process suggests that emotions are not discrete entities but rather emergent patterns arising from cognitive appraisals of situational meaning. The Tripartite Emotion Expression and Perception model \cite{SCHERER19-TEEP} builds on this perspective, proposing that the way humans perceive others' affective states is inherently linked to this appraisal process: individuals infer others' emotional states by unconsciously reconstructing the underlying appraisal process based on observable cues such as facial expressions, vocal prosody, and physiological reactions. From this perspective, appraisal dimensions provide a more robust foundation for affect recognition and generalisation across age groups than discrete affective labels.

The present study extends the THERADIA-WoZ corpus \cite{fournier2025theradia} 
to young adults using the same experimental design, enabling a direct cross-age 
comparison of affect expression, annotation, and recognition. A review of related work on affect recognition in AI-assisted healthcare, the theoretical and empirical comparison of representation types, and cross-corpus generalisation is provided in Section~\ref{sec:related_work}. Based on this review, the present study pursues three specific objectives:

\begin{enumerate}
    \item \textbf{Corpus extension and analysis.} A new corpus of young adults is introduced, following the same collection, segmentation, and annotation pipeline as the original THERADIA-WoZ corpus. A comprehensive analysis of the annotated labels and dimensions is presented, alongside a cross-corpus comparison leading to the identification of a refined core set of affective labels relevant across both age groups in the context of AI-assisted CCT.
    
    \item \textbf{Affect representation comparison.} Multimodal affect recognition models based on appraisal dimensions are compared against those based on categorical labels, within corpus and across corpora, using audio, visual, and textual modalities with both expert and deep learning representations.
    
    \item \textbf{Cross-age generalisation.} The extent to which models trained on one age group generalise to the other is evaluated under within-corpus, cross-corpus, and mixed-corpus training strategies, with the aim of determining whether appraisal dimensions provide a more stable representational framework for cross-age affect recognition than categorical labels.
\end{enumerate}

The remainder of this paper is organised as follows. Section~\ref{sec:related_work} reviews the relevant literature. Section~\ref{sec:corp_const} describes the data collection protocol. Section~\ref{sec: corp_a} presents the corpus analysis. Section~\ref{sec: model} details the automatic recognition experiments. Section~\ref{sec: gen} reports the generalisation results across age groups. Section~\ref{sec: conclusion} concludes and discusses the implications of the findings.

\section{Related Work}
\label{sec:related_work}

Building on the state of the art reviewed in the original THERADIA-WoZ study~\cite{fournier2025theradia} regarding appraisal theories, annotation methods, and existing healthcare corpora, this section focuses on three aspects central to the present extension: affect recognition in AI-assisted healthcare interactions, the empirical comparison of representation types, and cross-corpus generalisation across demographic groups.

\subsection{Affect Recognition in AI-Assisted Healthcare}

The integration of automatic affect recognition into healthcare applications has attracted growing attention, yet significant challenges remain. A systematic review of methods applied in real clinical contexts concluded that the field is characterised by significant heterogeneity in experimental protocols, small sample sizes, and a lack of robustness and reproducibility in reported results~\cite{pepa2021automatic}. Furthermore, models are 
frequently trained on one population and directly applied to another without cross-demographic validation, making it difficult to assess the reliability of obtained performances. This limitation is particularly consequential in contexts such as CCT, where the effectiveness of AI-assisted sessions is thought to depend on socially and emotionally adaptive interactions \cite{LAMPIT14-CCT}.

Efforts have begun to address this gap. Multimodal affect recognition integrated 
into socially assistive robots has been shown to be perceived as more engaging 
by older adults in care facilities compared to non-empathic versions~\cite{abdollahi_artificial_2023}, 
underscoring the clinical relevance of affect-aware AI agents for this population. 
Similarly, virtual coaching systems designed for healthy seniors have demonstrated 
the feasibility of multimodal emotion expression recognition in conversational 
human-machine interaction, with combined audio and video modalities outperforming 
unimodal approaches~\cite{palmero_exploring_2025}. Both approaches, however, 
rely on categorical annotation schemes, leaving open whether theoretically grounded 
representations such as appraisal dimensions would improve recognition performance 
and generalisability in these contexts.

\subsection{Categorical, Dimensional, and Appraisal-Based Representations}

The automatic recognition of affective states requires deciding how those states 
are to be represented. Categorical labels derived from discrete emotion theories 
offer a first approach~\cite{ekman1978manual}. While basic emotion labels (e.g., 
fear, disgust, happiness, sadness, and anger) are commonly used to model affective 
states, they are not necessarily the most appropriate in healthcare and AI-patient 
interaction contexts, as they refer to affect from an evolutionary perspective and 
fail to cover the achievement-related affective states most relevant in these 
settings~\cite{fournier2025theradia}.

Part of affect theorists suggest that labels, as social constructs, may fail to 
fully capture the underlying processes driving affective expression~\cite{BARRETT06SOL}. 
Instead, they propose that affect refers to changing states representable by 
dimensions, whose labels would try to capture certain spaces. The most popular 
dimensional theory proposes to differentiate affect based on valence, ranging from 
unpleasant to pleasant, and arousal, ranging from deactivation to 
activation~\cite{RUSSEL80-CIRCUM}. While this approach offers greater flexibility 
than categorical labels, it suffers from limitations as well. Certain labels such 
as anger and fear are placed in the same position in the circumplex, both being 
negative and aroused. In addition, valence is a multidimensional 
concept~\cite{FOURNIER22-COMB, fournier2024emotional}, and it has been proposed 
that two dimensions are not sufficient to differentiate affective 
states~\cite{FONTAINE07-DIM}.

Appraisal theories offer a more principled framework. They posit that affective states are causally determined by the cognitive evaluation of events along structured criteria \cite{SANDER13-MOD, SANDER05-CPM}, and that observable signals used in affect recognition are downstream manifestations of this appraisal process \cite{SCHERER19-TEEP}. This makes appraisal dimensions a more fundamental and theoretically stable annotation target than discrete labels. A survey of affective theory use in computational models of emotion confirmed that appraisal frameworks are increasingly adopted in affective computing precisely for this reason \cite{smith_what_2022}. In natural language processing, appraisal dimensions have been shown to constitute a reliable framework for emotion modelling in text, with appraisal patterns providing useful structure to represent and differentiate emotion categories~\cite{troiano_dimensional_2023}. Despite these theoretical 
advantages, the empirical comparison between appraisal-based and categorical 
representations in multimodal affect recognition, and in particular their relative 
generalisation capacity, has remained largely unexplored.

\subsection{Cross-Corpus Generalisation}

A central and persistent challenge in affective computing is the generalisation 
of trained models to unseen data distributions, attributed to differences in 
recording conditions, annotation schemes, speaker demographics, and language. 
Transfer learning approaches have been proposed to address this problem. 
Gerczuk et al.~\cite{gerczuk_emonet_2023} assembled a large multi-corpus dataset 
of 84,181 recordings from 26 SER corpora and proposed EmoNet, a residual adapter 
framework enabling parameter-efficient training of a shared model across corpora. 
The authors note, however, that only corpora with categorical labels were 
considered, leaving open the question of whether the approach extends to 
dimensional or appraisal-based annotation schemes. Latif et al.~\cite{latif_self_2023} 
proposed an adversarial dual discriminator network leveraging self-supervised 
pre-training to learn domain-invariant representations across five datasets and 
three languages without requiring target data labels, reducing the domain shift 
between training and test distributions. Taken together, these studies highlight 
that cross-corpus generalisation remains an open challenge, and that the role of 
the annotation scheme in mediating transfer capacity has yet to be systematically 
investigated.

Age constitutes a largely under-explored axis of domain shift. 
Poyiadzi et al.~\cite{poyiadzi_domain_2021} directly studied domain 
generalisation for facial expression recognition across age groups, 
finding that excluding age groups from training affects out-of-domain 
performance, particularly for neighbouring age groups, and that 
increasing the number of training age groups tends to improve 
generalisation to unseen ones. Their results, obtained on posed 
facial expressions, leave open whether appraisal-based representations 
might generalise more robustly across age. A systematic review of 
deep learning-based facial expression recognition for the 
elderly~\cite{GAYA25DEEP} further highlighted that existing datasets 
lack diversity in age representation and that real-world deployment 
of such systems remains scarce and experimental.

\subsection{Multimodal Affect Corpora and Dataset Positioning}
Large-scale corpora such as SEWA-DB~\cite{KOSSAIFI19-SEWA} have 
consolidated benchmarking practice in affective computing. However, none 
of these resources target AI-assisted CCT or include appraisal-based 
annotations. The few corpora from clinical or assistive settings 
\cite{palmero_exploring_2025, washington_improved_2022} use categorical 
or VAD annotations. No existing corpus enables a direct, controlled 
comparison of appraisal-based and categorical models across age groups 
within the same paradigm. 

The THERADIA-WoZ corpus \cite{fournier2025theradia} was the first multimodal affect corpus grounded entirely in appraisal theory and collected in an ecological healthcare interaction context. The present work extends this corpus to a young adult population, enabling for the first time a direct empirical evaluation of whether appraisal dimensions generalise better than categorical labels across age groups in AI-assisted CCT.

\section{Data Collection}
\label{sec:corp_const}
The protocol was reviewed and approved by the Ethics Committee for Research in Grenoble Alpes (CERGA-Avis-2021-1). Participants provided informed consent in accordance with the European General Data Protection Regulation (GDPR). The data collection protocol for this study follows the same methodology as the previously developed THERADIA-WoZ corpus \cite{fournier2025theradia}, with the difference that participants were young adults instead of older and MCI adults.

\subsection{Participants}
A total of 52 young adults (34 females; mean age = 21.17, SD age = 2.9) were recruited via flyers distributed on the campus of Lyon 2 University and through emails sent to students. Inclusion criteria included fluency in French, either normal or corrected-to-normal vision and hearing, and informed consent prior to recordings. In terms of educational attainment, 6\% of the sample possessed a master's degree, 31\% had obtained a bachelor's degree, 59\% had attained a baccalaureate degree, and 4\% held a National Vocational Qualification (French CAP) or a lower certification.

\subsection{Material}
The CCT session consisted of eight exercises selected from the HappyNeuronPro platform, targeting cognitive functions such as memory, language, attention, and planning. The virtual assistant was developed by Dynamixyz, and was controlled by a human experimenter using a Wizard-of-Oz (WoZ) paradigm (see \cite{fournier2025theradia} for details). The experimenter’s head movements, gaze, speech, and articulation were captured in real-time to animate the 3D virtual assistant displayed on the participant’s screen. The dialogue followed a predefined tree structure designed to accommodate a wide range of interactions, with provisions for free intervention when necessary.

The experimental setup mirrored that of the previous study. Sessions took place in quiet rooms equipped with a 24-inch monitor, a webcam (Logitech StreamCam), a headset with a unidirectional microphone (Sound BlasterX H6), and an external sound card (Presonus Audiobox). To ensure stable video recordings, an iPhone X was used alongside an omnidirectional microphone. Audio was sampled at 44.1 kHz (16-bit), and video was recorded at 30 fps (1920×1080 resolution, YUV420p format). Additional details regarding the hardware and software configurations can be found in \cite{fournier2025theradia}.

\subsection{Procedure}
Participants first completed consent forms and psychological, affective, and cognitive questionnaires before being divided into five groups (see \cite{ZSOLDOS-SUB-WOZ} for a detailed description and analysis of the questionnaires). Twenty participants, assigned to a non-induction group, completed two CCT sessions one week apart. The remaining 32 participants were placed in four induction groups, designed to elicit affectively charged expressions.

Each CCT session followed the same structure, consisting of eight exercises, each performed twice, with the difficulty level of the second attempt adjusted based on the participant’s initial performance. The session adhered to the five-phase format described in \cite{fournier2025theradia}, including introductory dialogues, exercise execution, feedback dialogues, repetitions, and a final debriefing.

For the induction groups, affect induction was applied in half of the eight exercises by manipulating three parameters: (i) presenting the exercise as easy or hard beforehand, (ii) setting the actual difficulty as easy or hard, and (iii) providing feedback from the WoZ that was either critical or congratulatory.

At the end of the last session, a debriefing was conducted to explain the cover story to participants.

For further details on the experimental protocol, task design, and affect induction strategies used in the original THERADIA-WoZ study, refer to \cite{fournier2025theradia}.

\subsection{Data segmentation and annotation}
The segmentation, transcription, and sequence selection followed the same methodology as the previous THERADIA-WoZ study \cite{fournier2025theradia}. Recordings were segmented based on pragmatic completeness in breath groups, ensuring semantic coherence, and transcribed using ELAN software following the ESLO project principles. Sequences containing affective expressions were retained, with redundant or excessive data removed to optimize corpus size.

The selection of affective labels and appraisal dimensions remained identical to the previous study. A total of 23 affective labels were used: angry, annoyed, anxious, ashamed, confident, contemptuous, curious, desperate, disappointed, embarrassed, excited, frustrated, interested, guilty, happy, hopeful, impatient, proud, relaxed, sad, satisfied, surprised, and upset. The five appraisal dimensions were: Novelty (predictable to unexpected), Intrinsic Pleasantness (unpleasant to pleasant), Goal Conduciveness (opposed to conducive with the participant’s wishes), Coping (unable to fully able to cope with the situation), and Arousal (asleep to awake).

Annotations were conducted using ANNOT, a web-based annotation tool. Six students from diverse academic backgrounds (engineering, business, and biology) were trained by project investigators, including an expert in appraisal theories, before conducting annotations remotely. Throughout the process, their work was monitored and supervised by the research team to ensure annotation consistency.

The dataset comprises approximately 30 hours of recorded data, segmented into 13540 transcribed sequences, whom 1496 were annotated (M = 8.37 s, SD = 4.62 s), totaling around 3 hours of annotated sequences, cf. table \ref{tab:corpus-stats}. 

\begin{table}[ht!]
\setlength{\tabcolsep}{13 pt}
\renewcommand{\arraystretch}{1.9}
\caption{Overall statistics of the data collected in the THERADIA-Woz. }

\label{tab:corpus-stats}
\begin{tabular}{lrrr}
\hline

Data & \multicolumn{1}{c}{\# Seq.} & \multicolumn{1}{c}{\begin{tabular}[c]{@{}c@{}}Seq. duration\\ Mean (std)\end{tabular}} & \multicolumn{1}{c}
{\begin{tabular}[c]{@{}c@{}}Overall \\ duration\end{tabular}} \\ 

\hline
Transcribed & 13540 &   7.81\,s (3.16\,s)   & 30h18min2s \\ 

\hline
Annotated   & 1496& 8.37\,s (4.62\,s)& 3h27m7.18s\\
		
\hline
\end{tabular}
\end{table}
\begin{figure*}[t!]
%\centering
\includegraphics[width=.95\textwidth]{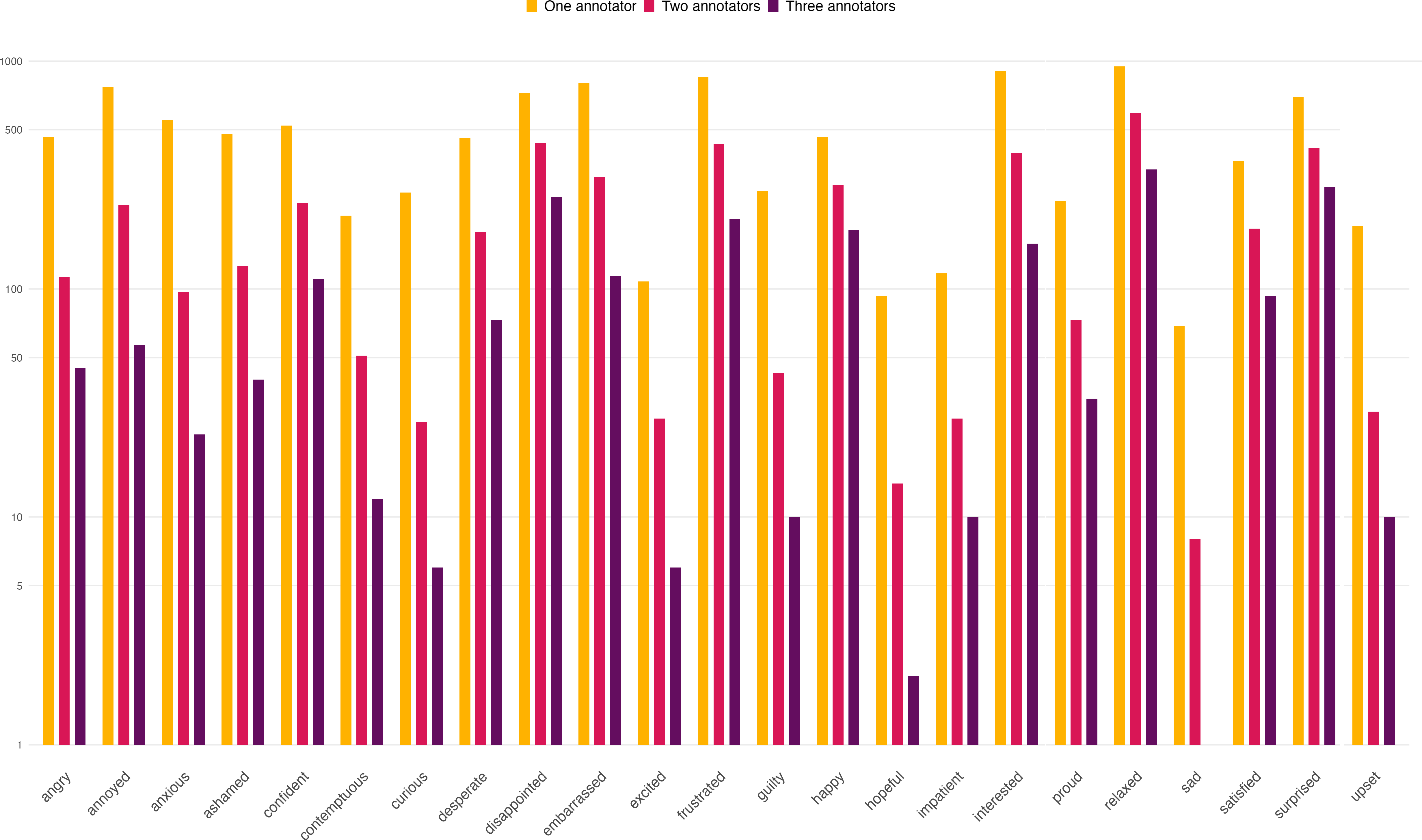}
\caption{Frequency (logarithmic scale) of the annotated labels according to the agreement of at least one, two or three annotators among six.}
\label{fig:freq_labels}
\end{figure*}

\section{Corpus Analysis}
\label{sec: corp_a}

%%% adding comparisons with old corpus

Building upon the methodological framework established in our previous study on older adults \cite{fournier2025theradia}, we followed a similar analysis approach to the young adults corpus. This includes presenting the annotation frequency of the affective labels as well as the inter-annotator agreement on their intensity. An analysis of inter-annotator agreement of dimensions and their relationships with affective labels was also evaluated. Finally, a comparison between the two corpora was performed to identify the most relevant affective labels in the context of AI-assisted CCT sessions. 
%This led to the identification of five key affective states. Namely, Annoyed, Frustrated, Interested, Relaxed and Surprised.

\subsection{Categorical annotations}
\label{sec:label_analysis}

For affective labels, we first examined their frequency by at least one, two, or three annotators, as illustrated in Figure~\ref{fig:freq_labels}. We computed Cohen’s $d$ to determine the minimum frequency threshold at which an effect size could be considered at least "small" (Cohen's $d$ > 0.2) ~\cite{cohen2016power}. For the young adults corpus, this threshold corresponded to $526$ sequences. Table \ref{tab:coreset-overview} presents the annotation frequency per label, with values exceeding this theoretical threshold highlighted in bold. Next, we assessed inter-annotator agreement on the presence of each affective label using the Unweighted Average Recall (UAR), considering instances where the label intensity exceeded zero. The results, detailed in Table \ref{tab:coreset-overview}, indicate that inter-annotator agreement exceeded the statistical threshold of 53\% according to a two-tailed $z$-test for all labels except for "sad". Those labels are marked in bold style in Table \ref{tab:coreset-overview}.

The inter-rater agreement on the intensity of the affective labels was evaluated using the Pearson's Correlation Coefficient (PCC). The inter-rater agreement was defined as sufficiently high if its Cohen's $d$, i.e., its PCC value divided by the sample standard deviation, is superior to .2~\cite{cohen2016power}, which corresponds to a PCC greater than $.024$. This concerned most of the affective labels, which are reported in bold style. 
%% comparison w OLd adults

%For each pair of annotators, we computed the Unweighted Average Recall (UAR) as follows:
%\begin{equation}
%UAR = \frac{1}{k} \sum_{i=1}^k \frac{N_c^i}{N^i}
%\label{UAR}
%\end{equation}
%where $k$ represents the total number of classes, which in this case corresponds to the presence or absence of a label. $N^i$ denotes the total number of annotated sequences for a given affect label, while $N_c^i$ represents the number of commonly identified labels, regardless of presence or absence. 

\begin{table}[t!]
\centering
\setlength{\tabcolsep}{15pt}
\renewcommand{\arraystretch}{1.6}
\footnotesize
\caption{Inter-annotator agreement on each sequence, or CCT session, computed with the Pearson's correlation coefficient for both time-continuous and summary values of appraisal dimensions; mean (standard-deviation).}
\label{tab:agreement_dim}

\begin{tabular}{lll}
\hline
& \multicolumn{1}{c}{Sequence} & \multicolumn{1}{c}{Session} \\ \hline
\multicolumn{3}{l}{Time-continuous values}                                          \\ \hline
Arousal&   .445 (.140)& .432 (.136)\\
Coping&   .506 (.081)& .379 (.143)\\
Goal conduciveness     &   .566 (.060)&  .475 (.132)\\
Intrinsic pleasantness &    .562 (.078)& .437 (.143)\\
Novelty&     .378 (.089)& .288 (.123)\\ \hline

\multicolumn{3}{l}{Summary values}                                                  \\ \hline
Arousal&   .437 (.118)& .330 (.220)\\
Coping&   .428 (.072)& .472 (.150)\\
Goal conduciveness     &   .490 (.059)& .562 (.162)\\
Intrinsic pleasantness &   .459 (.078)& .557 (.171)\\
Novelty&   .275 (.102)& .410 (.166)\\ \hline
\end{tabular}
\end{table}

\subsection{Dimensional annotations}
\label{dim-annot}

%The cognitive delay in the continuous annotation of each dimension was evaluated, and results showed that goal conduciveness (3.3\,seconds) had a significantly longer delay than the other dimensions (less than 2 seconds), suggesting that annotators had more difficulty assessing it. This evaluation is detailed in section III.A of the supplementary material.

We assessed the inter-annotator agreement on the dimensions for both time-continuous and summary annotations, using the PCC, which was computed either on each sequence, or each CCT session, cf. Table \ref{tab:agreement_dim}. 

For both time-continuous and summary annotations, goal conduciveness and intrinsic pleasantness show higher inter-annotator agreement consistently at both the sequence and session levels. In comparison, arousal and coping show lower agreement, with coping showing a relatively high correlation in time-continuous at the sequence level but dropping at the session level, whereas still consistent for summary annotations. Arousal follows a similar trend in summary values with moderate agreement at the sequence level, but sees a more pronounced drop in session-level. Novelty consistently demonstrates the lowest agreement across both time-continuous and summary annotations which was likely more difficult to annotate, as it is more context dependent in the study.

\begin{table}[t]
\centering
\setlength{\tabcolsep}{7pt}
\renewcommand{\arraystretch}{1.3}
\caption{Statistics on the annotated affective labels: frequency, inter-annotator agreement on the presence, and intensity, and correlation to appraisal's dimensions. Most relevant labels in the context of AI-assisted CCT are marked in bold}
\label{tab:coreset-overview}
\begin{tabular}{lrrrc}
\hline
\multicolumn{1}{c}{\multirow{2}{*}{Label}} &\multicolumn{1}{c}{\multirow{2}{*}{Frequency}} & \multicolumn{2}{c}{Agreement on}& \multirow{2}{*}{Correlation to} \\ \cline{3-4}
\multicolumn{1}{c}{}& \multicolumn{1}{c}{}& \multicolumn{1}{c}{Presence} &\multicolumn{1}{c}{Intensity} &\\ \hline

\multicolumn{5}{c}{\textbf{Core-set selection}} \\
\hline
\textit{Relaxed}& \textbf{950}& \textbf{61.37\%}& \textbf{216}& \cmark\\  
\textit{Interested}& \textbf{902}& \textbf{55.45\%}& \textbf{.105}& \cmark\\  
\textit{Frustrated}& \textbf{853}& \textbf{62.91\%}& \textbf{.284}& \cmark \\  
\textit{Annoyed}& \textbf{771}& \textbf{58.84\%}& \textbf{.207}& \cmark\\  
\textit{Surprised}& \textbf{694}& \textbf{65.79\%}& \textbf{.384}& \cmark\\ \hline

\multicolumn{5}{c}{\textbf{Not selected}} \\
\hline
\textit{Embarrassed}& \textbf{801}& \textbf{60.76\%}& \textbf{.209}& \cmark\\  
\textit{Disappointed}& \textbf{726}& \textbf{63.29\%}& \textbf{.267}& \cmark\\  
\textit{Anxious}& \textbf{551}& \textbf{54.89\%}& \textbf{.082}& \\  
\textit{Confident}& 522& \textbf{62.51\%}& \textbf{.268}&\\  
Ashamed& 479& \textbf{61.00\%}& \textbf{.208}&\cmark\\  
Angry& 464& \textbf{62.70\%}& \textbf{.283}&\cmark \\  
Happy& 465& \textbf{70.31\%}& \textbf{.444}&\cmark\\  
Desperate& 460& \textbf{60.98\%}& \textbf{.264}&\cmark\\  
\textit{Satisfied}& 364& \textbf{65.00\%}& \textbf{.357}&\cmark \\  
Guilty& 269& \textbf{54.91\%}& \textbf{.105}&\\  
Curious& 266& \textbf{55.53\%}& \textbf{.094}&\\  
Proud& 243& \textbf{64.87\%}& \textbf{.304}&\cmark\\  
Contemptuous& 210& \textbf{57.92\%}& \textbf{.217}&\\  
Upset& 189& \textbf{58.06\%}& \textbf{.176}&\cmark\\  
Impatient& 117& \textbf{58.46\%}& \textbf{.188}&\\  
Excited& 108& \textbf{56.85\%}& \textbf{.159}&\\  
Hopeful& 94& \textbf{55.63\%}& \textbf{.083}&\\  
Sad& 69& 51.80\%& \textbf{.041}&\\ \hline
\end{tabular}
\end{table}

\begin{figure}[t!]
\centering
\includegraphics[width=.47 \textwidth]{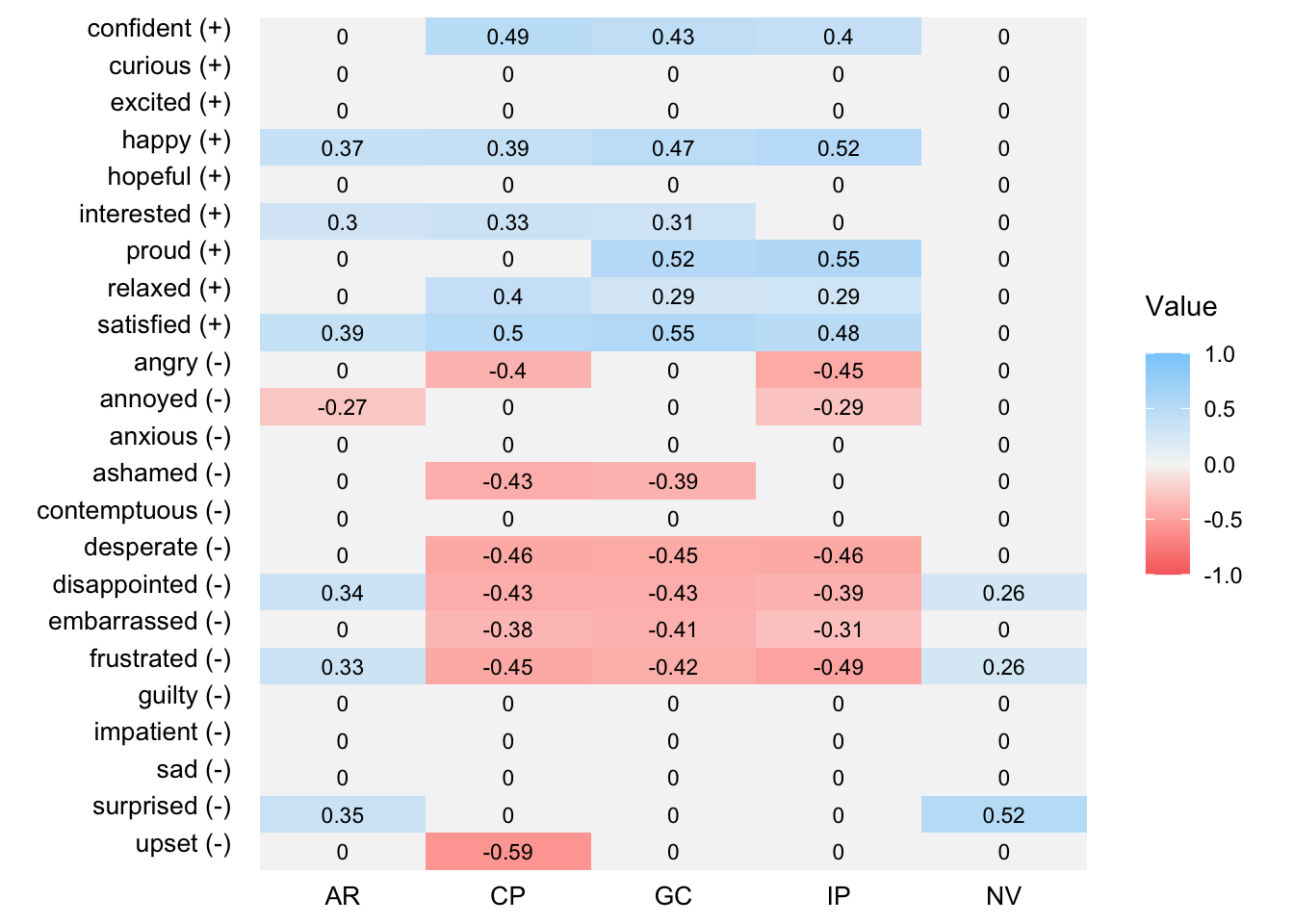}
\caption{Heatmap of the PCC between affective labels and summary values of appraisal dimensions aggregated across all annotators. The sign (+) or (-) after each label denotes whether it is positive or negative, respectively. Dimensions are as follows: AR: arousal, CP: coping, GC: goal conduciveness, IP: intrinsic pleasantness, NV: novelty. The correlations shown represent the average between annotators for whom the correlation was significant (\textit{p} < 0.05) after Bonferroni correction and of identical direction. Positive or negative direction is shown in blue or red, respectively. A correlation present for a single annotator was considered null.}

\label{fig_2}
\end{figure}

\subsection{Relation between dimensions and affective labels}
\label{sec:relation}

The relationship between appraisal dimensions and affective labels was investigated through the correlation patterns between summary values of appraisal dimensions and affective labels for each annotator separately. Specifically, Pearson's correlation coefficient (PCC) was computed for each dimension-label pair across sequences where the label was present. A correlation was deemed to be consistent if it was statistically significant and exhibited the same direction for a minimum of two annotators. The mean of these consistent correlations between annotators is presented in Figure \ref{fig_2}, and the labels that exhibited a significant relationship with at least one appraisal dimension are marked with "\cmark" in the Table \ref{tab:coreset-overview}.

The results are consistent with those reported in the original THERADIA-WoZ corpus. As anticipated, positive labels exhibit robust correlations with intrinsic pleasantness, goal conduciveness, and coping, while negative labels are inversely correlated with these dimensions. Additionally, surprise correlated with novelty, which aligns with theoretical expectations. These results are further corroborated by a meta-analysis on the relationship between appraisal dimensions and affective labels \cite{YEO24METACOR}.

Furthermore, our results emphasize the advantage of considering both intrinsic pleasantness and goal conduciveness, rather than valence alone, in differentiating affective states \cite{FOURNIER22-COMB}. For example, while both annoyed and desperate are negatively correlated with intrinsic pleasantness, only desperate is also negatively correlated with goal conduciveness. This distinction highlights the added value of appraisal dimensions over the valence/arousal model \cite{RUSSEL80-CIRCUM}, which often reduces valence to intrinsic pleasantness, neglecting its multidimensional nature that also encompasses goal conduciveness.

\subsection{Label core set selection}
In order to identify the core set of relevant affective labels, we applied the same selection process used in the previous analysis on the older adults corpus [6], considering four criteria defined a priori: (i) the frequency is significantly higher than the average on all affective labels; (ii) the agreement on presence is significantly higher than chance; (iii) the agreement on intensity is sufficiently high; and (iv) there is a significant correlation, in the same direction, with at least one dimension for at least two annotators. On the young adults corpus, this yielded seven candidate labels: Relaxed, Interested, Frustrated, Embarrassed, Annoyed, Disappointed, and Surprised, cf. Table \ref{tab:coreset-overview}. The cross-corpus comparison was then performed solely as a final intersection step, identifying the affective states consistently relevant across both age groups in the context of AI-assisted CCT, resulting in a final core set of Relaxed, Interested, Frustrated, Annoyed, and Surprised.

%A core set of the likely most relevant affective labels in the context of AI-assisted CCT was defined based on our analyses. %Results of these analyses are summarised in the Table \ref{tab:coreset-overview}. Four main criteria of selection were used: (i) the frequency is significantly higher than the average on all affective labels; (ii) the agreement on presence is significantly higher than chance; (iii) the agreement on intensity is sufficiently high; and (iv) there is a significant correlation, in the same direction, with at least one dimension for at least two annotators. Ten affective labels meet these four criteria, five positive and five negative, cf. Table \ref{tab:coreset-overview}. 

\section{Automatic recognition of affective states}
\label{sec: model}

%All experiments were conducted using 5fold cross validation.

\begin{figure}[t!]
\includegraphics[width=0.47\textwidth]{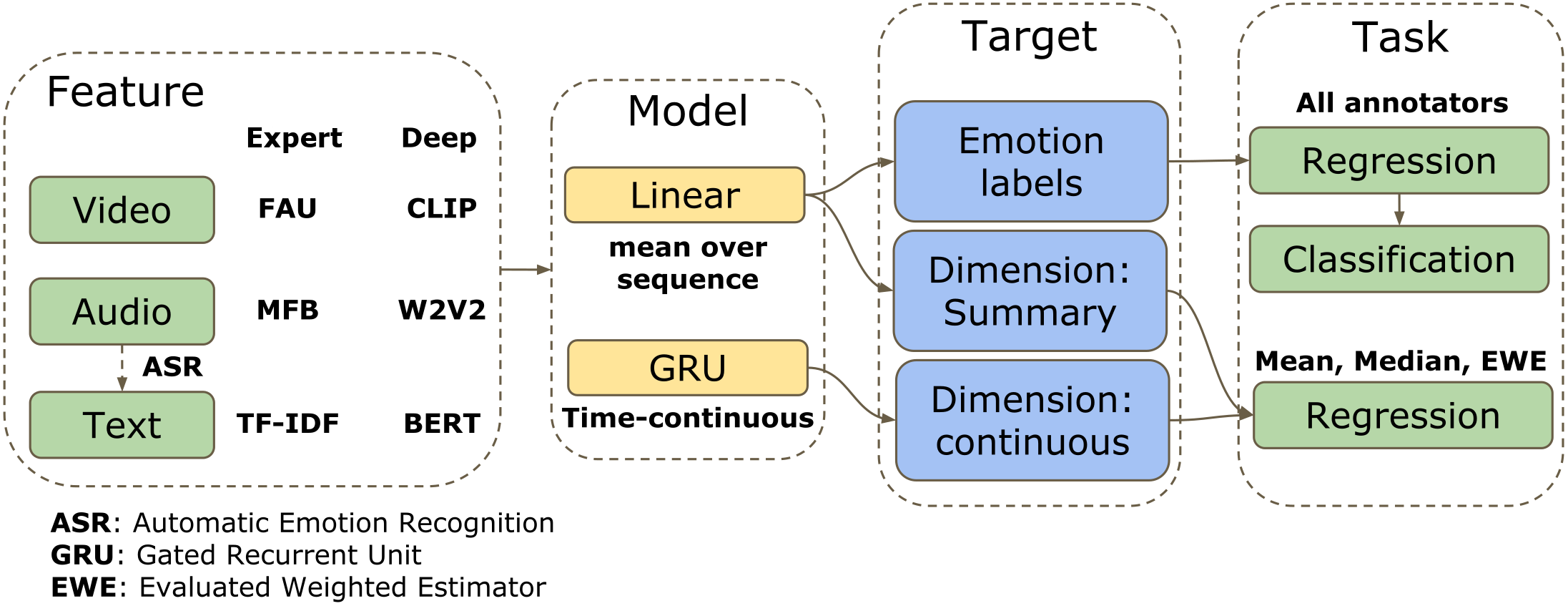}
\centering
\caption{Overview of the predictive experiments performed on the corpus. }
\label{fig:emo_rec_overview}
\end{figure}

In this section, we outline the approach used for the automatic prediction of affective labels and dimensions, maintaining the same framework as in our previous study to ensure consistency and comparability. The experimental setup is illustrated in Figure \ref{fig:emo_rec_overview}. We continue to use the same representations and models selected, extracting both expert and deep learning based representations from audio, textual, and visual data and training the same selected models for predicting affective labels, dimensions summaries and continuous dimensions.
For the text modality, we followed the same framework as in our previous work \cite{fournier2025theradia}, where the data were obtained through automatic speech transcription using Google ASR.

%\subsection{Representations of audio, textual and video data}
\subsection{Representations, Models, and Training Strategy}
\label{sec:rep}

In alignment with our previous methodology outlined in ~\cite{fournier2025theradia}, we employed expert-based features, including Mel-scale Filter Banks (MFBs)~\cite{davis1980comparison} for audio, Term Frequency - Inverse Document Frequency (TF-IDF)~\cite{salton1988term} for text, and Facial Action Units (FAUs)~\cite{ekman1978manual} for visual data. For deep representations, we used multilingual Wav2Vec2 model (trained on 56 languages, including French, and fine-tuned for ASR)
~\cite{baevski2020wav2vec} for audio, a multilingual BERT base model (fine-tuned for sentiment analysis on six languages, including French) for text ~\cite{devlin2018bert}, and CLIP~\cite{clip} for visual data. 

We used the gold standards that demonstrated the best performance in~\cite{fournier2025theradia}. For label intensity, models were trained to predict all six available annotations for each affective label. The average of these predictions was compared with the average of the six label targets for evaluation, following a methodology similar to~\cite{Ringeval15-POA}. For dimension summaries, the mean of the predictions was used. For continuous dimensions, we applied the Evaluator Weighted Estimator (EWE) technique ~\cite{stappen2021muse}.

For predicting label intensity and dimension summaries, we employed a Multi-Layer Perceptron (MLP) with one linear layer to map the features to the desired number of outputs, combined with a hidden linear layer of half the size of the input features. For continuous dimensions, we used a GRU model with three hidden layers of the same size as the input for deep features, while a hidden size of 256 cells was optimal for hand-crafted representations.

Multimodal fusion was achieved using a decision-based fusion approach. For label intensity and dimension summaries, we used an Ordinary Least Squares (OLS) regression model. For time-continuous dimensions, a GRU was employed as a weighted averaging layer to account for the temporal dynamics of the features, with weights optimised on the training set.

All models were trained using five-fold cross-validation, with folds constructed in a subject-independent manner within each corpus. The models were trained with the Adam optimizer and an initial learning rate of $10^{-3}$. The batch size and gradient accumulation were set to 10 for affective labels and dimensions summaries, and to 1 with a learning rate of $10^{-4}$ for continuous dimensions. The maximum number of training epochs was set to 50, with an early stopping strategy employed after five epochs. Training utilised the Mean Squared Error (MSE) as the loss function, and evaluation was performed using the Concordance Correlation Coefficient (CCC) on all sequences ~\cite{Weninger16-DTR}.

\subsection{Affect prediction for the core set labels}
\label{sec:affectIntensityPred}

The results for affective label prediction are presented in Table \ref{tab:label-regression}. A comparison between audio data captured through both an iPhone (farfield) and a headset (closetalk) as well as the comparison with multimodal models for older adults is provided in Table 1 of the supplementary material.

Compared to the older adults corpus, different patterns emerge, particularly in the dependencies between modalities and affective labels. Text-based features consistently outperform audio and video features across all labels. While audio features enhance predictions for Frustrated and Relaxed using both deep and expert representations, it shows lower performance than text modality, which can be due to variability in speech patterns among young individuals. Video expert features show weak performance, likely due to high intra-individual variability in facial expressions that the model fails to capture. However, CLIP-based features, which incorporate contextual information, improve the predictive performance particularly for the label Relaxed, whereas Interested remains challenging across both FAU and CLIP representations. The labels Surprised and Annoyed remain difficult to predict across all modalities. Among all modalities, text-based features offer a more robust representation than audio and video, which show challenges for the models in capturing consistent affective patterns.

Overall, the multimodal approach consistently outperforms unimodal models, with the fusion of unimodal models yielding more accurate predictions across all labels. However, when comparing the multimodal results to those from the older adults corpus, the latter demonstrates statistically superior performance.

%Similar patterns are observed for iPhone and Headset audio data, with this latter leading to better results for audio and text modalities due to its better audio quality, which enhances ASR performance. 
%comparing the performance of iPhone (farfield) and headset (closetalk) audio data, as well as the multimodal model trained and tested on the corpus of older adults.

\begin{table}[t]
\setlength{\tabcolsep}{8pt} % Adjusts column spacing to fit width
\renewcommand{\arraystretch}{1.53} % Adjusts row spacing
\caption{Core-set label intensity prediction (CCC) on young corpus as a function of Expert (TF-IDF, MFB, and FAU) and Deep (BERT, W2V2, CLIP) representations.}
\label{tab:label-regression}
\centering
\begin{tabular}{llcccc}
\hline
Label & Modality & Text & Audio & Video & Multimodal \\ \hline

\multirow{2}{*}{Annoyed}    & Expert & .103  & .042  & .043  & .148  \\  
& Deep   & .151  & .104  & .100  & .121  \\ \hline
\multirow{2}{*}{Frustrated} & Expert & .223  & .160  & .023  & .271  \\  
& Deep   & .228  & .169  & .093  & .270  \\ \hline
\multirow{2}{*}{Interested} & Expert & .135  & .080  & .021  & .171  \\  
& Deep   & .270  & .219  & .024  & .277  \\ \hline
\multirow{2}{*}{Relaxed} & Expert & .337  & .207  & .012  & .404  \\  
& Deep   & .407  & .287  & .287  & .400  \\ \hline
\multirow{2}{*}{Surprised}  & Expert & .093  & .092  & .046  & .126  \\  
& Deep   & .178  & .128  & .131  & .194  \\ \hline
\multirow{2}{*}{Average}    & Expert & .202  & .116  & .033  & .224  \\  
& Deep   & .247  & .181  & .127  & .252  \\ \hline
\end{tabular}
\end{table}

\subsection{Dimensions summaries prediction}
\label{dim_sum_Sec}

The results for the five dimensions prediction are shown in Table \ref{tab:dim-sum}. Additional comparisons, including the evaluation of multimodal models for older adults and the comparison of the performance between farfield (iPhone) and closetalk (headset) audio, are presented in Table 2 of the supplementary material.

The text modality consistently achieves the highest performance among all modalities for predicting the five dimensions, offering a more robust representation. Arousal and Novelty show the lowest predictive performance, making them more challenging compared to other dimensions. This aligns with the correlation analysis \ref{fig_2}, which shows their correlation with the labels Annoyed and Surprised, where similar patterns were observed in affective label prediction. Audio features contribute to improved predictions for Coping, Goal Conductiveness, and Intrinsic Pleasantness. However, as seen with labels prediction, FAU-based features yield low performance, whereas incorporating contextual information through CLIP significantly enhances predictions for these three dimensions.

The multimodal approach consistently yields higher performance across all dimensions. When comparing age groups, younger adults show lower predictive performance across most dimensions, especially for Coping, Goal Conduciveness, and Intrinsic Pleasantness. This difference can likely be explained by the high variability in the emotional expressions of younger adults, whereas older adults show more stable and consistent affective patterns making their affective states less challenging to model and predict.

\begin{table}[t!]
\setlength{\tabcolsep}{2pt}
\renewcommand{\arraystretch}{1.6}
\scriptsize
\centering
\caption{Dimensions summaries prediction (CCC) on young corpus as a function of Expert (TF-IDF, MFB, and FAU) and Deep (BERT, W2V2, CLIP) representations.}
\label{tab:dim-sum}
\resizebox{\linewidth}{!}{
\begin{tabular}{llcccc}
\hline
Dimension & Modality & Text& Audio  & Video  & Multimodal \\ \hline
\multirow{2}{*}{Arousal}                & Expert & .193 & .098 & .007 & .215 \\  
                                        & Deep   & .111 & .076 & .059 & .207 \\ \hline
\multirow{2}{*}{Coping}                 & Expert & .400 & .260 & .053 & .440 \\  
                                        & Deep   & .369 & .226 & .239 & .474 \\ \hline
\multirow{2}{*}{Goal Conduciveness}     & Expert & .412 & .205 & -.017 & .450 \\  
                                        & Deep   & .286 & .249 & .253 & .473 \\ \hline
\multirow{2}{*}{Intrinsic Pleasantness} & Expert & .368 & .295 & .057 & .433 \\  
                                        & Deep   & .346 & .184 & .308 & .480 \\ \hline
\multirow{2}{*}{Novelty}                & Expert & .224 & .029 & .018 & .267 \\  
                                        & Deep   & .201 & .173 & .021 & .234 \\ \hline
\multirow{2}{*}{Average}                & Expert & .319 & .178 & .024 & .361 \\  
                                        & Deep   & .262 & .182 & .176 & .374 \\ \hline
\end{tabular}
}
\end{table}

\subsection{Continuous dimensions prediction}

The results for the continuous dimensions are presented in Table \ref{tab:dim-cont}. The evaluation of multimodal models in older adults and the comparison between farfield (iPhone) and closetalk (headset) audio are provided in Table 3 of the supplementary material. These results are consistent with the dimensional summary analyses, particularly in demonstrating the robustness of the textual representation.  Furthermore, the multimodal approach consistently outperforms unimodal models across all dimensions, with superior performance systematically observed in older adults.
% statistical tests
Among the dimensions, Arousal is better predicted across all three modalities in the continuous case, likely due to its more stable temporal variations, making it easier to model dynamically. In contrast, Goal Conduciveness and Coping show low predictive performance with video features. Novelty remains challenging dimension to predict, consistent with previous findings, due to its  context dependent nature, which makes it a challenging dimension to predict.

\begin{table}[t!]
\setlength{\tabcolsep}{2pt}
\renewcommand{\arraystretch}{1.6}
\scriptsize
\centering
\caption{Continuous dimension prediction (CCC) on young corpus as a function of Expert (TF-IDF, MFB, and FAU) and Deep (BERT, W2V2, CLIP) representations.}
\label{tab:dim-cont}
\resizebox{\linewidth}{!}{
\begin{tabular}{llcccc}
\hline
Dimension & Modality & Text & Audio & Video & Multimodal \\ \hline
\multirow{2}{*}{Arousal}                & Expert & NA  & .251 & .199 & NA  \\  
                                        & Deep   & .363 & .329 & .278 & .406 \\ \hline
\multirow{2}{*}{Coping}                 & Expert & NA & .095 & .056 & NA \\  
                                        & Deep   & .319 & .267 & .169 & .348 \\ \hline
\multirow{2}{*}{Goal Conduciveness}     & Expert & NA & .165 & .052 & NA \\  
                                        & Deep   & .417 & .286 & .126 & .386 \\ \hline
\multirow{2}{*}{Intrinsic Pleasantness} & Expert & NA & .145 & .119 & NA \\  
                                        & Deep   & .348 & .231 & .183 & .290 \\ \hline
\multirow{2}{*}{Novelty}                & Expert & NA & .057 & .060 & NA \\  
                                        & Deep   & .205 & .146 & .070 & .178 \\ \hline
\multirow{2}{*}{Average}                & Expert & NA & .142 & .097 & NA \\  
                                        & Deep   & .330 & .251 & .165 & .322 \\ \hline
\end{tabular}
}
\end{table}

\section{Comparison between annotation type and Generalisation across age}
\label{sec: gen}
\subsection{Overview of Hypotheses and Bayesian analysis plan}

This section investigates whether multimodal models based on time-continuous and summary dimensions outperform those based on labels in affect recognition. Additionally, we assess the extent to which these differences persist across age groups by evaluating generalisation effects and whether training on both age groups could enhance generalisation. To address these hypotheses, three training strategies were applied: (1) within-corpus evaluation, where models were trained and tested on the same age corpus; (2) cross-corpus evaluation, where models were trained on one age corpus and tested on the other; and (3) mixed-corpus evaluation, where models were trained on a corpus combining both age corpora and tested separately on each age corpus. All analyses used CCC as the evaluation metric. Given that results from expert representations were not available for time-continuous dimensions, only results from models using deep representations are reported in the main text. The results of experiments conducted using expert representations for labels and summary dimensions are provided in the supplementary material and revealed similar patterns to those obtained with deep representations.

To statistically test these effects, a Bayesian linear model was fitted to Fisher z-transformed CCC values (z-CCC), incorporating main effects and interaction effects for affect representation, training strategy, and the age corpus used as the test set. Bayesian inference was used to provide direct evidence for or against hypotheses, specifically testing whether an effect impacts the data or not \cite{KRUSCHKE15}. Bayes Factors (BF) were reported to quantify evidence, with BF$_{10}$ indicating evidence in favor of the alternative hypothesis and BF$_{01}$ supporting the null hypothesis . Regarding BF interpretation, a BF $\geq$ 3 indicates moderate evidence, a BF $\geq$ 10, suggests strong evidence, while a BF $\geq$ 100 is considered decisive \cite{KASS95BF}. Planned contrasts were performed to assess effects identified as having an impact on data with multiple modalities. This was done using the Region of Practical Equivalence (ROPE) framework to determine whether performance differed from an interval considered null based on the data type. As commonly applied in Bayesian inference, ROPE was defined as half of a small effect \cite{KRUSCHKE13}, specifically 0.1 in the context of coefficient of correlation \cite{COHEN13COR}. Default priors were used in all Bayesian models, with four Markov Chain Monte Carlo chains of 40,000 iterations each.

\subsection{Results}

\begin{figure*}[t!]
%\centering
\includegraphics[width=.95\textwidth]{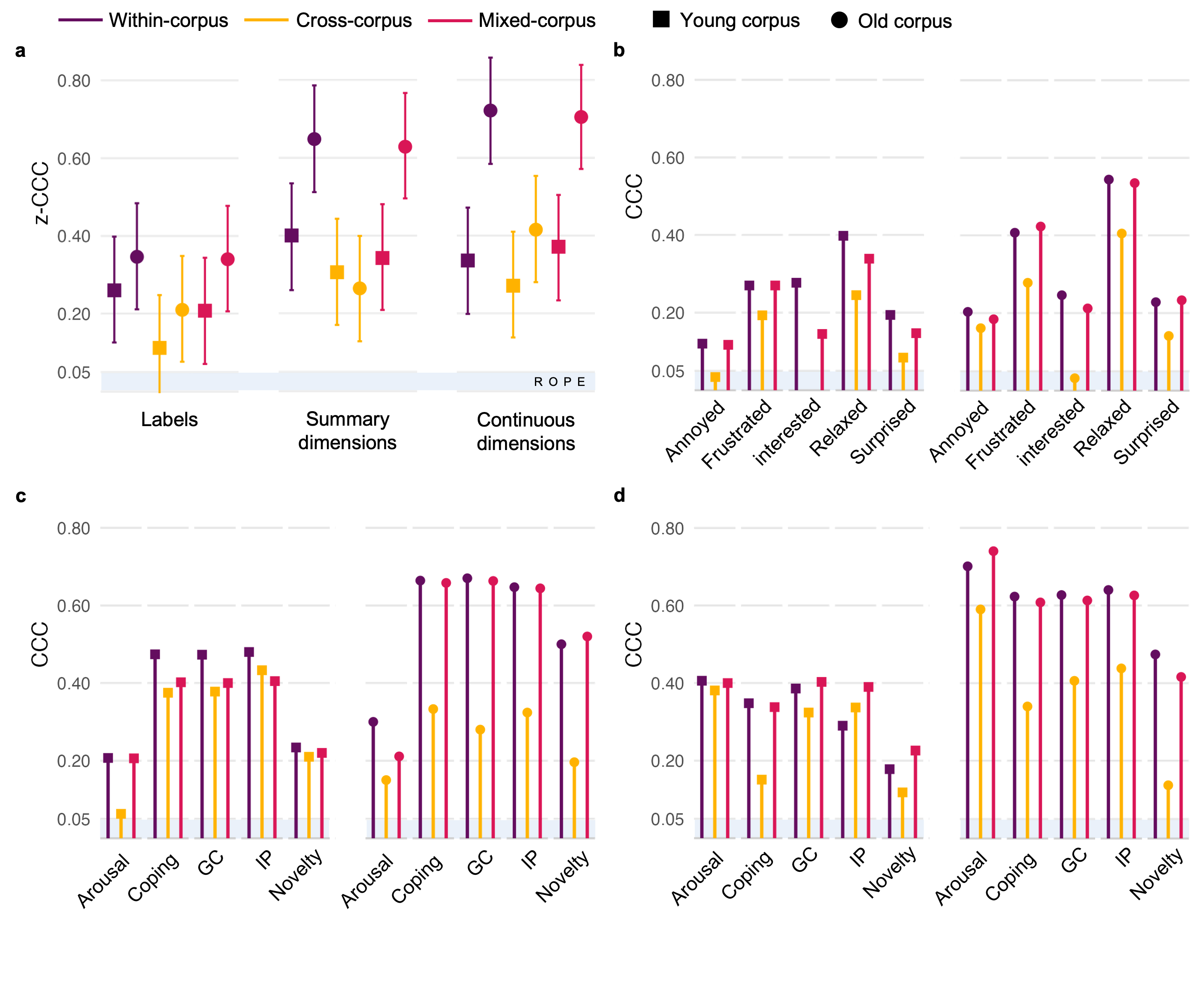}
\caption{Results of multimodal predictions using deep representations across affect representation (Labels, Summary Dimensions, Time-Continuous Dimensions), test corpora (Young vs. Older Adults), and training strategies (Within-Corpus, Cross-Corpus, Mixed-Corpus). In each plot, blue area represents the Region of Practical Equivalence (ROPE) defined as half of a small effect (0.1). \textbf{a}. Medians and 95\% credible intervals of Fisher z-transformed CCC values by modality, extracted from model's posterior distributions. \textbf{b}. CCC for each core-set label. \textbf{c}. CCC of each summary dimension. \textbf{d}. CCC for each time-continuous dimension. \textbf{c,d}. GC refers to Goal Conduciveness; IP refers to Intrinsic Pleasantness.}

\label{fig:sum_gen}
\end{figure*}

\subsubsection{Assessing Performance Against Chance}
Before comparing model performance across affect representations, training strategies, and test corpora, we first examined whether predictions were meaningfully above chance using ROPE contrasts.

For labels, within-corpus performance provided moderate evidence of superiority over the ROPE when tested on the young adult corpus (BF$_{10}$ = 7) and decisive evidence when tested on the older adult corpus (BF$_{10}$ > 100). This pattern was relatively well distributed across labels, with consistently higher CCC values for \textit{Relaxed} in both corpora and lower CCC values for \textit{Annoyed}. However, in cross-corpus evaluation, label performance fell within the ROPE when tested on the young adult corpus (BF$_{01}$ = 19) and remained inconclusive when tested on the older adult corpus (BF$_{10}$ = 2). The decline in performance in cross-corpus settings was consistent across labels, but specific labels such as \textit{Interested} in both corpora and \textit{Frustrated} when tested on the young adult corpus had CCC values that fell within the ROPE, lowering the overall effect. In mixed-corpus evaluation, performance remained inconclusive when tested on the young adult corpus (BF$_{10}$ = 2) but was decisively above ROPE when tested on the older adult corpus (BF$_{10}$ > 100), suggesting that while performance remained stable, slight reductions from the within-corpus condition caused some labels to fall within the ROPE.

For summary dimensions, within-corpus performance showed decisive evidence of superiority over ROPE in both young and older adults (BF$_{10}$ > 1,000). Performance was consistently strong for \textit{Coping}, \textit{Intrinsic Pleasantness}, and \textit{Goal Conduciveness}, while lower for \textit{Arousal} in both corpora and \textit{Novelty} when tested on the young adult corpus. In cross-corpus evaluation, strong evidence of superiority was found in both young adults (BF$_{10}$ = 57) and older adults (BF$_{10}$ = 14), maintaining the same pattern observed in within-corpus evaluations. Similarly, in mixed-corpus training, performance remained decisively above ROPE for both young and older adults (BF$_{10}$ > 100 and BF$_{10}$ > 1,000, respectively), indicating stability across conditions.

For time-continuous dimensions, within-corpus evaluation demonstrated decisive evidence of superiority over ROPE when tested on the young adult corpus (BF$_{10}$ > 100) and older adults (BF$_{10}$ > 1,000). The patterns were similar between the two corpora, though \textit{Novelty} was slightly lower, while \textit{Arousal} showed better performance when tested on the older adult corpus. In cross-corpus evaluation, performance remained strongly or decisively above ROPE (BF$_{10}$ = 23 and BF$_{10}$ = 14, respectively), with a more pronounced drop for \textit{Coping} when tested on the young adult corpus. In mixed-corpus training, performance remained decisively superior across both test corpora (BF$_{10}$ > 1,000), following the same trends observed in within-corpus evaluations.

These results indicate that performance was meaningfully above chance for both dimensional representations and labels in within-corpus settings, though labels struggled to maintain performance above the ROPE in cross-corpus conditions. Time-continuous and summary dimensions consistently showed stronger evidence of robustness across all training strategies and test corpora.

\subsubsection{Comparison of performance across conditions}

Regarding affect representation, models trained on time-continuous and summary dimensions consistently outperformed those trained on labels. As shown in Figure \ref{fig:sum_gen}a, z-CCC values were higher for both dimensional representations across all training strategies and test corpora. Bayesian analysis confirmed a significant main effect of affect representation (BF$_{10}$ > 1,000), with both time-continuous and summary dimensions yielding better predictions than labels (BF$_{10}$ > 1,000 and BF$_{10}$ > 100, respectively). Equivalence was found between time-continuous and summary dimensions (BF$_{01}$ = 18).

Regarding test corpus, models performed better when tested on the older adult corpus compared to the young adult corpus. As illustrated in Figure \ref{fig:sum_gen}a, z-CCC values were consistently higher for the older adult test set across all affect representations and training strategies. Bayesian analysis supported this observation, revealing a significant main effect of test corpus (BF$_{10}$ > 1,000), confirming superior model performance when applied to the older adult corpus.

Regarding training strategy, within-corpus and mixed-corpus training resulted in better predictive performance than cross-corpus training. Figure \ref{fig:sum_gen}a shows a marked decline in z-CCC values for cross-corpus training, particularly for categorical labels. Bayesian analysis revealed a significant main effect of training strategy (BF$_{10}$ > 100), with both within-corpus (BF$_{10}$ > 100) and mixed-corpus training (BF$_{10}$ = 10) outperforming cross-corpus training. No significant difference was found between within-corpus and mixed-corpus training (BF$_{01}$ = 25).

The pattern of results for these three main effects appears stable across their interactions. Bayesian analysis supported this, showing no interaction effects between affect representation and training strategy (BF$_{01} >$ 1,000), affect representation and test corpus (BF$_{01} = 26$), and training strategy and test corpus (BF$_{01} = 15$). Additionally, the absence of a triple interaction among all three factors was supported (BF$_{01}$ > 1,000).

Taken together, these results indicate that dimensional annotations consistently outperform categorical labels, that models perform better when tested on older adults, and that within-corpus and mixed-corpus training yield better generalisation than cross-corpus training. Notably, categorical labels failed to generalise across age corpora, as performance dropped to chance level in cross-corpus training. In contrast, dimensional representations maintained predictive performance above chance, highlighting their robustness for multimodal affect recognition across age groups. However, training on both corpora (mixed-corpus) did not further enhance generalisation, as performance remained similar to within-corpus training, suggesting that exposure to both age groups did not provide additional benefits for model adaptation.
\section{Conclusion} % Sina: removed "and discussion" since we did not discuss anything in the end
\label{sec: conclusion}
Recently, the THERADIA-WoZ corpus was developed, allowing the scientific community to train models on multimodal affect recognition in a healthcare context, namely, AI-assisted Computerized Cognitive Training. The THERADIA-WoZ corpus provided two types of affect recognition, relying both on categorical labels and dimensions borrowed from appraisal theories of emotion, and was based on an older adults age group of participants. Here, we introduce an extension of the THERADIA-WoZ corpus based on young adult participants, enabling a direct comparison of affect recognition models across age groups. The purpose of this study was to assess whether multimodal models based on appraisal dimensions outperform those based on categorical labels and to evaluate the extent to which these differences persist across age corpora. We hypothesized that appraisal dimensions would offer superior predictive performance and generalisation compared to labels, which was confirmed by our findings.

The identification of a core-set of labels, derived from the five most robust labels across both the young and older adult corpora, consists of a key contribution of this work. This selection is particularly meaningful in the context of AI-assisted CCT, as it provides a refined set of affective categories that are most relevant for this application. By consolidating the core-set based on empirical evidence from two corpora, we contribute to a more targeted and effective representation of affect in AI-driven healthcare interactions.

Furthermore, the analysis conducted on the relationship between appraisal dimensions and affective labels is in alignment with the findings from meta-analyses on this subject \cite{YEO24METACOR}. Moreover, the results emphasize the benefits of appraisal dimensions in the context of affect differentiation. First, the findings underscore the benefit of incorporating both intrinsic pleasantness and goal conduciveness in the differentiation of affective states, as opposed to solely considering valence. For instance, while both Annoyed and Desperate exhibit a negative correlation with intrinsic pleasantness, only Desperate demonstrates a concurrent negative correlation with goal conduciveness. This lends further support to the argument that appraisal dimensions provide a more nuanced representation of affect in comparison to the common bidimensional valence/arousal model \cite{RUSSEL80-CIRCUM}, which summarises valence as intrinsic pleasantness and neglects its multidimensional nature.

Our study also confirms previous findings from THERADIA-WoZ regarding multimodal fusion, which consistently outperforms unimodal representations from text, audio, or facial expressions alone. This reinforces the idea that emotional expression is inherently multimodal, and using multiple sources of information improves the ability to recognise emotions. Similarly, as previously observed in THERADIA-WoZ, deep representations provide superior predictions compared to expert features, underscoring the effectiveness of learned representations in capturing emotional nuances.

With respect to our hypotheses, the results revealed that appraisal dimensions consistently outperform categorical labels. Notably, categorical labels failed to generalise across age corpora, as performance dropped to chance level in cross-corpus training. This finding aligns with existing literature indicating that emotion expression differs across age groups \cite{GAYA25DEEP, FOLSTER14EXP, KO21CHANGE, SONMEZ19COMP, ATALLAH19REV, JANNAT21EXP, PARK22FAC, AMORIM21CHAN, LIN24AGE, SEN18AGE, ANDY22TWIT}. In contrast, predictions from appraisal maintained performance above chance, highlighting their robustness for multimodal affect recognition across age groups. This suggests that appraisal dimensions provide a more stable and generalisable framework for affect recognition, effectively overcoming the limitations observed with categorical labels. 

The results also showed overall better performance when models were tested on the older adult corpus compared to the young adult corpus. One possible explanation is that older adults may have been more engaged in the CCT sessions, leading to a richer range of emotional expressions, which in turn facilitated more effective training. This interpretation is supported by the questionnaire-based analyses conducted on the same dataset within the THERADIA project \cite{ZSOLDOS-SUB-WOZ}, which showed that older adults reported higher engagement, greater appreciation of the virtual assistant, less desire to give up, and lower fatigue at the end of the session compared to young adults. On the other hand, training on both corpora did not improve performance beyond within-age group predictions, but it also did not degrade it.

In conclusion, this study highlights the superiority of appraisal dimensions over categorical labels in multimodal affect recognition, particularly in their ability to generalise across age groups. Our findings indicate that dimensions derived from appraisal theories offer a more stable and theoretically grounded framework for emotion modelling while also exhibiting greater robustness in cross-age generalisation. These insights contribute to the multidisciplinary field of affective sciences. 

%%Additionally, we provide an API for researchers interested in predicting time-continuous emotion recognition (both labels and dimensions) based on the models trained in this study. This resource aims to support behavioural sciences by enabling more refined and data-driven measurements of individuals’ emotional states.

\section{Acknowledgments}
This research has received funding from the Banque Publique d’Investissement (BPI) under grant agreement THERADIA, the Association Nationale de la Recherche et de la Technologie (ANRT), under grant agreement No. 2019/0729, and has been partially supported by MIAI@Grenoble-Alpes, (ANR-19-P3IA-0003). Artificial intelligence (ChatGPT, OpenAI) was used solely for English language editing. All scientific content, interpretations, and conclusions are the responsibility of the authors.

%\balance

\bibliographystyle{IEEEtran}
\bibliography{refs.bib}

    %\includepdf[pages=-]{sup.pdf}

\IEEEpubid{0000--0000/00\$00.00~\copyright~2021 IEEE}

\vfill

\end{document}